\documentclass{appolb}
\usepackage{graphicx}
\usepackage{amsmath}
\usepackage{amsfonts}
\usepackage{amssymb}
\usepackage{physics}
\usepackage{cite}

\newcommand{\Yt}{{\mathbf{Y}}}

\newcommand{\rt}{{\mathbf{r}}}
\newcommand{\bt}{{\mathbf{b}}}
\newcommand{\xt}{{\mathbf{x}}}

\newcommand{\Deltat}{{\boldsymbol{\Delta}}}
\newcommand{\as}{\alpha_{\text{s}}}
\newcommand{\qs}{Q_{\text{s}}}

\newcommand{\nc}{N_\text{c}}

\newcommand{\nr}[1]{(\ref{#1})} 
\newcommand{\ud}{\, \textrm{d}}

\newcommand{\xpom}{{x_{\mathbb{P}}}}
\newcommand{\xbj}{{x_\text{Bj}}}

\newcommand{\conz}{\conj{0}}
\newcommand{\cono}{\conj{1}}

\newcommand{\contrip}{\conj{0} \conj{1} \conj{2}}
\newcommand{\conj}[1]{\mathop{\overline{#1}}\nolimits}

\newcommand{\cxt}{{\conj{\xt}}}

\begin{document}
\title{Diffractive structure function in the dipole picture%
\thanks{
Presented   at ``Diffraction and Low-$x$ 2022,'' Corigliano Calabro (Italy), September
24–30, 2022
} 
}
\author{G. Beuf
\address{National Centre for Nuclear Research, 02-093 Warsaw, Poland}
\\[3mm]
{H. H\"anninen, T. Lappi, H. M\"antysaari 
\address{Department of Physics, %
 P.O. Box 35, 40014 University of Jyv\"askyl\"a, Finland and 
Helsinki Institute of Physics, P.O. Box 64, 00014 University of Helsinki,
Finland}
}
\\[3mm]
Y. Mulian
\address{Instituto Galego de Física de Altas Enerxías IGFAE,
Universidade de Santiago de Compostela, 15782 Santiago de Compostela, Galicia-Spain}
}

\maketitle

\begin{abstract}
We calculate the contribution from the $q\bar{q}g$ component of a virtual photon state to the small-$x$ diffractive cross section in deep inelastic scattering in the saturation regime. The obtained cross section is finite by itself and a part of the full next-to-leading order result. We perform the calculation in exact kinematics in the eikonal limit, and show that the previously known high virtuality $Q^2$ and large invariant mass $M_X^2$ results for the structure functions can be extracted. We furthermore discuss the steps required to obtain the full next-to-leading order result.
\end{abstract}

\section{Introduction}

At high collision energy or small $x$, Deep Inelastic Scattering can be conveniently described in the dipole picture, combined with an eikonal scattering approximation for the scattering of the dipole off the target. 
When gluon saturation is important, the number of gluons given by a gluon distribution is not the best way to quantify the strength of the gluonic field of a hadronic target. 
In the high collision energy limit, the interaction of a high-energy probe  should be eikonal, meaning that the transverse coordinate of the probe does not change during the scattering. The basic degrees of freedom in our description are thus eikonal scattering amplitudes. 

The amplitude for the simplest colored probe, a single quark, has a microscopical interpretation in terms of a path-ordered exponential, Wilson line $V(\xt)$.
The Wilson line also provides a direct connection to the classical field in the initial stage of a heavy ion collision. 
In the case of DIS, the relevant dilute probe is a color-neutral quark-antiquark dipole. Its scattering amplitude is given by the dipole amplitude 
\begin{equation}
\mathcal{N}_{01} = 1-S_{01}= 1-\left< 
 \frac{1}{\nc}
\tr V(\xt_0)
V^\dag(\xt_1)\right>,
\end{equation}
which automatically interpolates between color transparency at $|\xt_{01}| \equiv |\xt_0-\xt_1|=0$ and saturation at $|\xt_{01}|\gtrsim 1/\qs$. 

This picture leads directly to the \emph{dipole picture of DIS}. Here, at leading order, the DIS scattering is factorized into the $\gamma^*$ fluctuating into a $q\bar{q}$ dipole, described by the photon light-cone wave function,  and the dipole amplitude. The total cross section is given, through the optical theorem, with the dipole then transforming back to a virtual photon. Diffractive DIS, at the focus of our attention  in Ref.~\cite{Beuf:2022kyp}, requires a specific color neutral final state after the target color field. At leading order the final state is another $q\bar{q}$ dipole, while  at NLO it includes an additional $q\bar{q}g$ state. 

\section{Diffractive structure function at leading order}

The diffractive structure function $ F_2^{D(3)}(\beta, Q^2,\xpom)$ (or equivalently the diffractive $\gamma^*$-target cross  section     $\ud \sigma^{\text{D}}_{\lambda,  \, q \Bar q}/\ud M_X^{2} \ud \abs{t}$, see \cite{Marquet:2007nf} for more details) is   a function of three Lorentz-invariant kinematical variables. In addition to the conventional $Q^2$ and $\xbj=\beta\xpom$, it also depends on how the  $\gamma^*$-target energy is divided between the diffractive system $X$ and the rapidity gap, with $\beta = Q^2/(Q^2+M_X^2),$ where $M_X$ is the invariant mass of the diffractive system. In this paper we are interested in the regime where $\beta$ is not parametrically small (i.e. $M_X$ not parametrically large), so that powers of $\as \ln 1/\beta$ do not need to be resummed. In different regimes of $0<\beta<1$, the cross section is dominated by different kinds of partonic configurations of the virtual photon~\cite{Golec-Biernat:1999qor}. At $\beta\to 1$, one predominantly has a $q\bar{q}$ state in a longitudinal total helicity configuration, and at $\beta\approx 1/2$ in a transverse one. Although these LO $q\bar{q}$ contributions have been known already, in  Ref.~\cite{Beuf:2022kyp} we derived new expressions allowing for a completely general impact parameter dependence of the cross section:
\begin{multline}
    \frac{\ud \sigma^{\text{D}}_{\lambda,  \, q \Bar q}}{\ud M_X^{2} \ud \abs{t}}
    = 
    \frac{\nc}{4\pi} \int_{0}^1 \!\! \ud z
 \!\!\! \!\!\!\!\!\!     \int\limits_{ \xt_0 \xt_1  \Bar \xt_0  \Bar\xt_1}
 \!\!\!\!\!\!   {\cal I}_{\Deltat}^{(2)} {\cal I}_{M_X}^{(2)}
   \widetilde{\psi}_{\gamma^{*}_\lambda \rightarrow q_{\Bar 0} \Bar{q}_{\Bar 1}}^\dagger
    \widetilde{\psi}_{\gamma^{*}_\lambda \rightarrow q_0 \Bar{q}_1}
    \big[S_{\conz \cono}^\dagger -1 \big]\big[S_{01} -1 \big] ,
\label{eq:loresult}
\end{multline}
where $z$ is a longitudinal momentum fraction.
The result is expressed in terms of ``transfer functions'' ${\cal I}_\Deltat^{(2)}$, 
${\cal I}_{M_X}^{(2)}$, relating the $(q,\bar{q})$ coordinates  in the amplitude ($\xt_0,\xt_1$) and conjugate ($\xt_{\conj{0}}, \xt_{\conj{1}}$) to $t$ and $M_X$ as
\begin{eqnarray}
    {\cal I}_\Deltat^{(2)}
    &=& 
=
\frac{1}{4\pi} J_0\left( \sqrt{\abs{t}}\: \norm{
z \xt_{\Bar 0 0} - (1-z)\xt_{\Bar 1 1}
} \right),
\\
 {\cal I}_{M_X}^{(2)}
   & =&
=
\frac{1}{4\pi} J_0\left( \sqrt{z (1-z)}M_X \norm{\Bar \rt-\rt} \right).
\end{eqnarray}
At $\beta \ll 1$, one starts to become sensitive to higher invariant mass states in the photon, meaning Fock states with a higher number of partons. The first one of these is the $q\bar{q}g$ state. While the $q\bar{q}g$ Fock component is in general a NLO correction, in the small $\beta$ regime it is in fact the leading contribution. Thus contributions where a gluon is emitted before the target, and survives until the final state, are a special class of NLO corrections that are meaningful to consider separately.

\section{Contribution of the $q\bar{q}g$ Fock state}

\begin{figure}
\centerline{\includegraphics[width=0.2\textwidth]{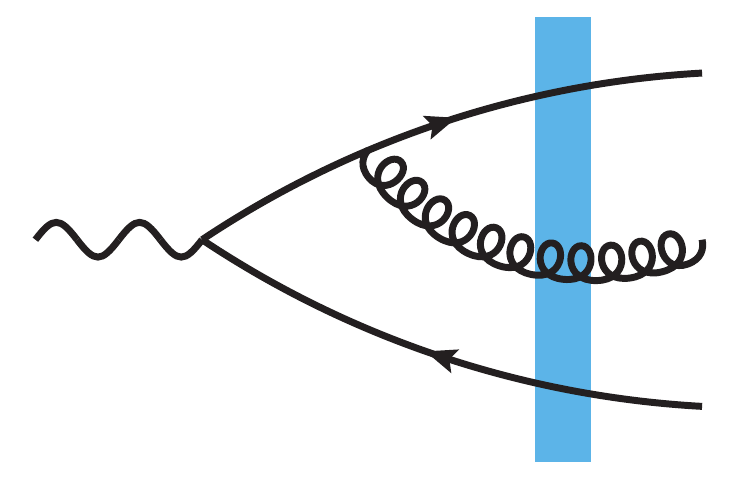}
\includegraphics[width=0.2\textwidth]{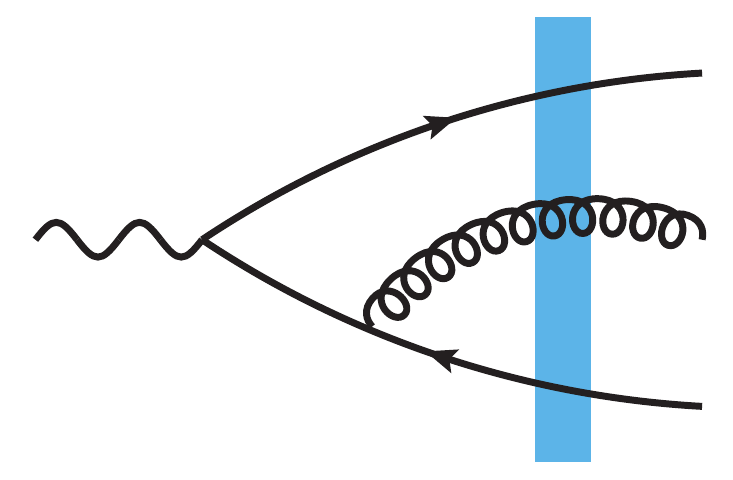}
\includegraphics[width=0.2\textwidth]{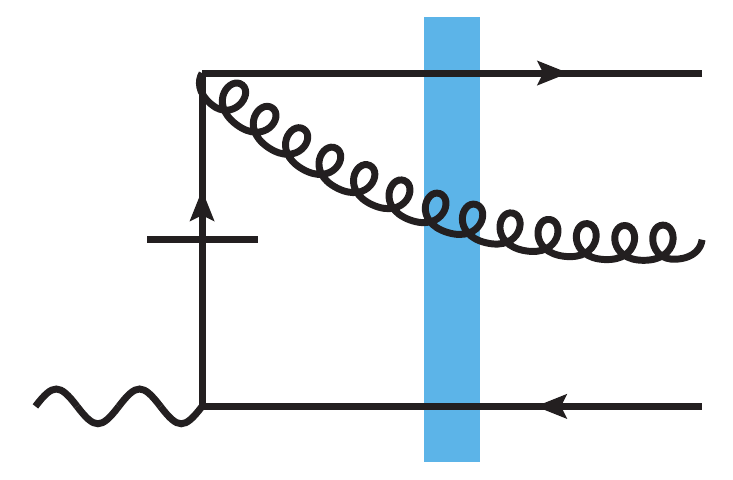}
\includegraphics[width=0.2\textwidth]{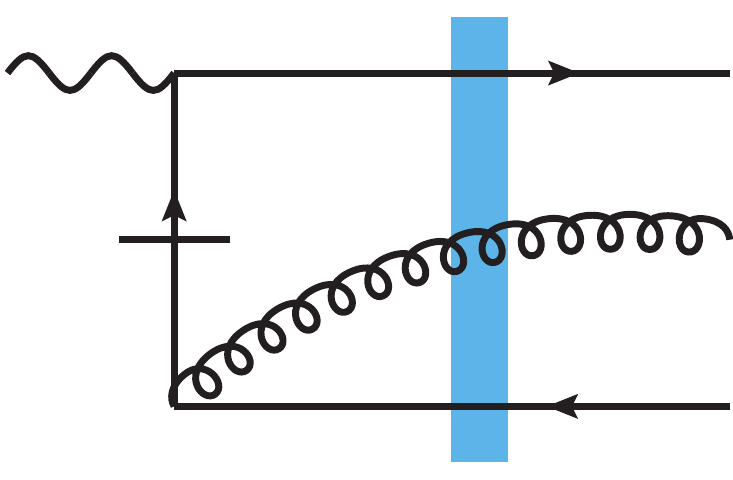}
}
\caption{Contributions to the amplitude where a gluon is emitted before the target, and produced and measured in the final state.}
\label{fig:nloraddiags}
\end{figure}

In Ref.~\cite{Beuf:2022kyp} (see also \cite{Hanninen:2021byo}), we calculated the full contribution of the diagrams shown in Fig.~\ref{fig:nloraddiags} to the diffractive structure function. The calculation is performed in what we refer to as ``exact eikonal kinematics,'' 
where the only kinematical approximation is the eikonal interaction with the target. The results are expressed, analoguously to Eq.~\nr{eq:loresult}, in terms of ``transfer functions'' ${\cal I}_{M_X}^{(3)}$ and ${\cal I}_{\Deltat}^{(3)}$.
They are then multiplied by the squared gluon emission wavefunction:
\begin{multline}
\xpom F_{T,L, \, q \Bar q  g}^{\textrm{D}(4) \, \textrm{NLO}} (\xbj, Q^2, \beta, t)
=
\!\!	\int_0^1 \!\! \frac{\ud z_0}{z_0}
	\frac{\ud z_1}{z_1}
	\frac{\ud z_2}{z_2}
	\delta(z_0 \! + \! z_1 \! + \! z_2 \! - \! 1)
\int_{\xt_0,\xt_1,\xt_2,\cxt_0,\cxt_1,\cxt_2}
\\
\times		{\cal I}_{M_X}^{(3)} {\cal I}_{\Deltat}^{(3)}\ 
  \widetilde{\psi}_{\gamma^{*}_\lambda \rightarrow q_{\Bar 0} \Bar q_{\Bar 1} g_{\Bar 2}}^\dagger
    \widetilde{\psi}_{\gamma^{*}_\lambda \rightarrow q_0 \Bar{q}_1 g_2}
	\left[ 1 - S_{\contrip}^{(3)\dagger} \right] \left[ 1 - S_{012}^{(3)} \right] ,
\end{multline}
where the final factors represent the ``tripole'' Wilson line operators for the $q\bar{q}g$ states  to interact with the target~\cite{Beuf:2017bpd}.

The squared wavefunctions are lengthy, but straightforward expressions in terms of the transverse coordinates and momentum fractions; they can be found in Ref.~\cite{Beuf:2022kyp}. It is interesting to discuss the different scales appearing in the expressions. The photon wavefunctions involve, just as in the leading order case, Bessel $K$ functions that exponentially cuf off configurations that are too large compared to the photon virtuality $Q$. The argument of the Bessel function is the square root of 
\begin{equation}
			Q^2X_{012}^2 
			 =
			Q^2\left[z_{0} z_{1} \xt_{01}^2 + z_{0} z_{2} \xt_{02}^2 + z_{1} z_{2} \xt_{12}^2\right],
\end{equation}
interpreted as the ratio of the formation time of the $q\bar{q}g$ system with coordinates $\xt_{0},\xt_{1},\xt_{2}$ (with $ \xt_{01} \equiv\xt_{0}-\xt_{1}$ etc.) to the lifetime of the virtual photon.  The 3-particle transfer function to $M_X$ is 
\begin{equation}  
  {\cal I}_{M_X}^{(3)} 
        = 
        2 \frac{z_0 z_1 z_2}{(4 \pi)^2} \frac{M_X}{Y_{012}} \ \mathrm{J}_1 \!\left( M_X Y_{012} \right),
\end{equation}
with  the coordinate combination conjugate to $M_X$
\begin{equation}  
    \Yt_{012}^2 =
        z_0 z_1 \left( \xt_{\Bar 0 0} - \xt_{\Bar 1 1} \right)^2 +
        z_1 z_2 \left( \xt_{\Bar 2 2} - \xt_{\Bar 1 1} \right)^2 +
        z_0 z_2 \left( \xt_{\Bar 2 2} - \xt_{\Bar 0 0} \right)^2.
\end{equation}
Finally, $t$ is conjugate to the  3-particle center-of-mass coordinate
\begin{equation}
    {\cal I}_\Deltat^{(3)} = \frac{1}{4 \pi} \  \mathrm{J}_0 \! \left( \sqrt{-t} \norm{z_0 \xt_{\Bar 0 0} + z_1 \xt_{\Bar 1 1} + z_2 \xt_{\Bar 2 2}} \right).
\end{equation}

\section{Towards a full NLO computation}

There are several additional contributions that will still need to be included for a full NLO result for the diffractive structure function. From these additional contributions it is not easy to define any further subsets that would be finite by themselves, but they will all have to be considered together in future work. Firstly, in addition to the ``emission before target'' radiative corrections in Fig.~\ref{fig:nloraddiags}, there are the corresponding gluon emissions from quarks after the target. Unlike the contributions that we have included, however, they will have a collinear divergence from the gluons being emitted at a small transverse momentum with respect to the emitting (anti)quark. These collinear divergences are cancelled by wavefunction renormalization of the outgoing quarks, i.e. by ``propagator correction'' diagrams on the outgoing quark lines. These in turn also include a UV divergence, and thus they have to be considered together with virtual contributions. There are several such virtual corrections. The most straightforward of these are the loop corrections to the $\gamma^* \to q\bar{q}$ wavefunction, which have already been calculated in Refs.~\cite{Beuf:2016wdz,Hanninen:2017ddy}. There are also contributions where the gluon is emitted before the target, and then crosses the target but is reabsorbed without being measured, analoguously to the $q\bar{q}g$ contributions in~\cite{Beuf:2017bpd,Hanninen:2017ddy}. They  involve both UV divergences that must be subtracted and canceled with the ones from before and after the target, and large logarithms of longitudinal momenta that must be factorized into BK/JIMWLK evolution of the target. Finally, there are final state interactions with gluon exchanges between the quark and antiquark after the target. These are a rather novel kind of contribution in this context and there is still some discussion about the correct way of treating them. 

\section{Known limiting cases}

While our calculation of the diffractive structure function in the ``exact eikonal kinematics'' is new, we have also checked that it reduces in specific limiting cases to results that are already available in the literature. In the limit of large $M_X$ we obtain the result (see \cite{Munier:2003zb} and references therein)
\begin{equation}
    \xpom F_{T,q\Bar q g}^{\textrm{D},(\text{MS})}
    = 
    \frac{\as \nc C_\text{F} Q^2}{16\pi^5\alpha_{\mathrm{em}}} 
\!\!\!
\!\!\!
    \int\limits_{ \xt_0 \xt_1 \xt_2} 
\!\!\!\!\!\!\!\!\!
\int_0^1 
\!\!\!
\ud z \frac{    |\widetilde{\psi}^{\rm LO} |^2 }{z(1-z)} 
 \frac{\xt_{01}^2}{\xt_{02}^2\xt_{12}^2} 
  \bigg[ \mathcal{N}_{02} + \mathcal{N}_{12} - \mathcal{N}_{01} - \mathcal{N}_{02}\mathcal{N}_{12} \bigg]^2 ,
\end{equation}
which factorizes into the LO $\gamma^*\to q\bar{q}$ wavefunction, the BK kernel for the emission of a soft gluon, and a squared Wilson line operator. This result is straightforwardly obtained from ours by first approximating the gluon as being soft $z_2\to 0$, in which case $M_X$ becomes dominated by the light cone energy of the gluon. One then removes the constraint on $M_X$ by integrating over the momentum fraction of the gluon $z_2$. This unconstrained integration introduces a divergence, which is however cured if final state emissions are included in the same kinematical approximation.

A rather more nontrivial task is to rederive the ``Wüsthoff result'' \cite{Wusthoff:1997fz,Golec-Biernat:1999qor} in the limit of large $Q^2$: 
\begin{multline}
\xpom  F_{T,q\Bar q g}^{\textrm{D},(\text{GBW})}  = \frac{\as\beta}{8\pi^4}\sum_{f}e_f^2 
\int_{\bt}
\int_{\beta}^{1} \!\! \ud z 
 \int_0^{Q^2} \ud k^2 k^4 \ln \frac{Q^2}{k^2}
 \\  \times
 \left[\int_0^\infty \ud r 
 r  K_2(\sqrt{z}k r) J_2(\sqrt{1-z}kr)  \mathcal{N}_\text{adj}(\bt,\rt,\xpom) \right]^2.
\label{eq:whoff}
\end{multline}
This result is characterized by an explicit logarithmic dependence on $Q^2$, and a $g\to q\bar{q}$ splitting function that can be associated with \emph{target DGLAP evolution}. It also depends on an adjoint representation dipole operator, with the small-size octet $q\bar{q}$ pair after the gluon emission acting as an effective gluon. The wavefunction of the photon splitting into this effective gluon dipole has a rank-2 traceless structure, which results in Bessel functions $J_2$ and $K_2$ in the final result. The key point in reaching this result starting from the dipole picture is to recognize that $z$ and $\beta$ in Eq.~\nr{eq:whoff} are to be understood as target ($k^-$) momentum fractions. To arrive at this form it is convenient to identify the target momentum fraction variables using invariant masses of the $q\bar{q}$ and $q\bar{q}g$ states before and after the scattering. One then  finds the large $Q^2$ limit by looking at the aligned jet kinematical limit for the dipole picture, with $z_0\ll z_1\ll z_2$. The derivation of the Wüsthoff result in the literature has not been very clearly documented, and certainly the original approach used to derive it has been very different than the one we use here. It is encouraging that as a side result our work has resulted in an independent rederivation of this widely and successfully used expression.

\subsubsection*{Acknowledgements}
{

\footnotesize
T.L and H.M are supported by the Academy of Finland, the Centre of Excellence in Quark Matter (project 346324) and projects 338263, 346567 and 321840. G.B is supported in part by the National Science Centre (Poland) under the research grant no. 2020/38/E/ST2/00122 (SONATA BIS 10). Y.M acknowledges financial support from Xunta de Galicia (Centro singular de investigación de Galicia accreditation 2019-2022); the ``María de Maeztu'' Units of
Excellence program MDM2016-0692 and the Spanish Research State Agency under project PID2020-119632GB-I00; European Union ERDF. G.B and Y.M acknowledge financial support from MSCA RISE 823947 ``Heavy ion collisions: collectivity and precision in saturation physics'' (HIEIC). This work was also supported under the European Union’s Horizon 2020 research and innovation programme by the European Research Council (ERC, grant agreement No. ERC-2018-ADG-835105 YoctoLHC) and by the STRONG-2020 project (grant agreement No. 824093). The content of this article does not reflect the official opinion of the European Union and responsibility for the information and
views expressed therein lies entirely with the authors.

}

\bibliographystyle{h-physrev4mod2}
\bibliography{difflowx22_lappi.bib}

\end{document}